\documentclass[a4paper,fleqn,usenatbib]{mnras}
\usepackage{newtxtext,newtxmath}
\usepackage[T1]{fontenc}
\usepackage{ae,aecompl}

\makeatletter
\def\@printed{}
\def\@journal{}
\def\@oddfoot{}
\def\@evenfoot{}
\makeatother

\usepackage{xfrac}
\usepackage{graphicx}
\usepackage{dcolumn}
\usepackage{bm}
\usepackage{hyperref}
\usepackage{xspace}
\usepackage{soul}
\usepackage[ruled,vlined]{algorithm2e}
\usepackage{dsfont}

\newcommand{\NATBIBORDER}[1]{}

\usepackage{aas_macros}

\newcommand{\code}{\textsf}
\newcommand{\inlinecode}{\texttt}

\newcommand{\nlive}{n_\text{live}}

\newcommand{\ndynamic}{n_\text{dynamic}}

\newcommand{\Z}{\mathcal{Z}}

\newcommand{\like}{\mathcal{L}}
\newcommand{\threshold}{\like^\star}
\newcommand{\pg}[2]{p\mathopen{}\left(#1\,\rvert\, #2\right)\mathclose{}}
\newcommand{\Pg}[2]{P\mathopen{}\left(#1\,\rvert\, #2\right)\mathclose{}}
\newcommand{\p}[1]{p\mathopen{}\left(#1\right)\mathclose{}}
\newcommand{\intd}{\text{d}}
\newcommand{\params}{\mathbf{\Theta}}

\newcommand{\expectation}[1]{\langle #1 \rangle}
\newcommand{\plateaulike}{\ensuremath{\like_P}}

\DeclareMathOperator{\image}{Im}
\DeclareMathOperator{\size}{size}
\DeclareMathOperator{\BinomDist}{Binom}
\DeclareMathOperator{\BetaDist}{Beta}

\newcommand{\uniform}{\mathcal{U}}

\usepackage{cleveref}

\usepackage[usenames]{xcolor}

\newcommand{\MN}{\code{MultiNest}\xspace}
\newcommand{\PC}{\code{PolyChord}\xspace}

\newcommand{\anesthetic}{\code{anesthetic}}
\newcommand{\version}{\code{\href{https://github.com/williamjameshandley/anesthetic/releases/tag/2.0.0-beta.2}{2.0.0-beta.2}}}

\title{Nested sampling with plateaus}

\author[A. Fowlie et al.]{%
    Andrew Fowlie$^{1}$\thanks{andrew.j.fowlie@njnu.edu.cn},
    Will Handley$^{2,3}$\thanks{wh260@cam.ac.uk},
    and Liangliang Su$^{1}$\thanks{liangliangsu@njnu.edu.cn}
    \\
$^{1}$Department of Physics and Institute of Theoretical Physics, Nanjing Normal University, Nanjing, Jiangsu 210023, China\\
$^{2}$Astrophysics Group, Cavendish Laboratory, J.J.Thomson Avenue, Cambridge, CB3 0HE, UK\\
$^{3}$Kavli Institute for Cosmology, Madingley Road, Cambridge, CB3 0HA, UK
}

\date{}
\pubyear{2020}

\begin{document}
\label{firstpage}
\pagerange{\pageref{firstpage}--\pageref{lastpage}}
\maketitle

\begin{abstract}
It was recently emphasised by \citet{riley,2020arXiv200508602S} that in the presence of plateaus in the likelihood function nested sampling (NS) produces faulty estimates of the evidence and posterior densities. After informally explaining the cause of the problem, we present a modified version of NS that handles plateaus and can be applied retrospectively to NS runs from popular NS software using \anesthetic. In the modified NS, live points in a plateau are evicted one by one without replacement, with ordinary NS compression of the prior volume after each eviction but taking into account the dynamic number of live points. The live points are replenished once all points in the plateau are removed. We demonstrate it on a number of examples. Since the modification is simple, we propose that it becomes the canonical version of Skilling's NS algorithm.
\end{abstract}

\begin{keywords}
    methods: statistical -- methods: data analysis -- methods: numerical
\end{keywords}

\section{Introduction}

Nested sampling (NS)~\citep{2004AIPC..735..395S,Skilling:2006gxv} is a popular algorithm for Bayesian inference in cosmology, astrophysics and particle physics. The algorithm handles multimodal and degenerate problems, and returns weighted samples for parameter inference as well as an estimate of the Bayesian evidence for model comparison. It was recently emphasised independently by \citet[appendix B.5.4]{riley} and \citet{2020arXiv200508602S}, however, that assumptions in NS are violated by plateaus in the likelihood, that is regions of parameter space that share the same likelihood. We should not be surprised by subtleties caused by ties in the likelihood as NS is based on order statistics and this problem and possible solutions were in fact discussed by \citet{2004AIPC..735..395S,Skilling:2006gxv}. We believe, however, that it was underappreciated prior to \citet{riley,2020arXiv200508602S}.

Plateaus most often occur in discrete parameter spaces in which every possible likelihood value is associated with a finite prior mass, or in physics problems when regions of a model's parameter space make predictions that are in such severe disagreement with observations that they are assigned a likelihood of zero. For example, in particle physics, parameter points that fail to predict electroweak symmetry breaking would be vetoed and in cosmology portions of parameter space may be excluded if they result in unphysically large power spectra for the purposes of applying lensing, or if it is impossible to trace a consistent cosmic history (see section XIII.C of \citealt{Hergt:2020} for more detail). We  generically refer to such points as unphysical and the observation that renders them unphysical as $U$. In fact, within popular implementations of NS, such as  \MN~\citep{Feroz:2007kg,Feroz:2008xx,Feroz:2013hea} and \PC~\citep{Handley:2015fda,Handley:2015xxx}, unphysical points can be assigned a likelihood of zero or a prior of zero through the \inlinecode{logZero} setting. Any log likelihoods below it are treated as if the prior were in fact zero. Statistically, this makes a difference to the evidences and Bayes factors, as it changes whether we consider $U$ to be prior knowledge or information with which we are updating, i.e.\ whether we wish to compute
$\pg{D}{M, U}$ or $\pg{D, U}{M}$,
where $D$ represents experimental data and $M$ represents a model. The latter is problematic within NS. We consider both cases interesting: on the one hand, taking a basic observation, e.g., electroweak symmetry breaking, as prior knowledge is reasonable, but, on the other, so is judging models by their ability to predict a basic observation. Although the problem with plateaus can in general lead to faulty posterior distributions as well, when plateaus occur only at zero likelihood, they do not impact posterior inferences about the parameters of the model. There are, furthermore, realistic situations in which plateaus could occur at non-zero likelihood, e.g., if in some regions of parameter space, the likelihood function or the physical observables on which it depends are approximated by a constant.

In \citet{riley,2020arXiv200508602S}, the problem caused by plateaus was formally demonstrated. After reviewing the relevant aspects of NS in \cref{sec:ns}, in \cref{sec:plateaus}, we instead make an informal explanation of the problem. We continue in \cref{sec:modified_ns} by proposing a modified NS algorithm that deals with plateaus and reduces to ordinary NS in their absence. We show examples in \cref{sec:examples} before concluding in \cref{sec:conclusions}. An implementation of the modified NS algorithm that can be used to correct evidences and posteriors found from \MN and \PC runs is implemented in \code{anesthetic} starting from version \version~\citep{Handley:2019mfs}.

\section{Nested sampling}\label{sec:ns}

To establish our notation and the assumptions in ordinary NS, let us briefly review the NS algorithm. NS works by computing evidence integrals,
\begin{equation}\label{eq:Z}
\Z \equiv \int_{\Omega_\params}  \like(\params) \, \pi(\params) \,\intd \params,  
\end{equation}
where $\like(\params)$ is the so-called likelihood function for the relevant experimental data and $\pi(\params)$ is the prior density for the model's parameters,
as Riemann sums of a one-dimensional integral,
\begin{equation}
\Z = \int_0^1 \like(X) \,\intd X,
\end{equation}
where 
\begin{equation}\label{eq:X}
X(\lambda) = \int_{\like(\params) > \lambda}  \pi(\params) \,\intd \params,
\end{equation}
is the prior volume contained within the iso-likelihood contour at $\lambda$ and $\like(X)$ is the inverse of $X(\lambda)$, i.e. $\like(X(\lambda)) = \lambda$. This evidently requires that such an inverse exists over the range of integration.

To tackle the one-dimensional integral, we first sample $\nlive$ points from the prior --- the live points. Then, at each iteration of NS, we remove the live point with the worst likelihood $\threshold$ and replace it with one drawn form the prior subject to the constraint that $\like > \threshold$. Thus we remove a sequence of samples of increasing likelihood $\like_i$. In NS we estimate $X_i \equiv X(\like_i)$ by the properties of how that sequence was generated. Indeed, at each iteration the volume contracts by a factor $t$, where the arithmetic and geometric expectations of $t$ are
\begin{align}\label{eq:t}
    \expectation{t} &= \frac{\nlive}{\nlive + 1},\\
    \expectation{\ln t} &= -\frac{1}{\nlive}.
\end{align}
We can then make a statistical estimate of the prior volume at the $i$-th iteration, $X_i = e^{-i/\nlive}$. This enables us to compute
\begin{equation}
\Z \approx \sum \like_i (X_{i - 1} - X_{i}).
\end{equation}
The estimates in \cref{eq:t} assume that the live points are uniformly distributed in $X$. In \citet{10.1093/mnras/staa2345} we proposed a technique for testing the veracity of this assumption within the context of a numerical implementation such as \code{MultiNest} or \code{PolyChord}. In the following section we discuss why plateaus violate that assumption.

\section{Plateaus}\label{sec:plateaus}

In \citet{Skilling:2006gxv}, plateaus were recognised as a problem, since they offer no guidance about the existence or location of points of greater likelihood and since they cause ambiguity in the ranking of samples by likelihood. The latter problem was addressed by breaking ties by assigning a rankable second property to each live point, that is expected to be unique, such as a cryptographic hash of the parameter coordinates or just a random variate from some pre-specified distribution. This was implemented by extending the likelihood to include that tie-breaking second property, $\ell$, suppressed by a tiny factor, $\epsilon$, so that it doesn't impact the evidence estimate,
\begin{equation}\label{eq:extened_like}
    \like \to \like + \epsilon \ell.
\end{equation}
It was not stated explicitly that plateaus violate an assumption in NS, as formally shown by \citet{riley,2020arXiv200508602S}, though it was known and it was mentioned in \citet[section 4.4.6]{murray}. \citet{murray}, furthermore, noted that hashes of the parameter coordinates in discrete parameter spaces would not be unique and thus would fail to break ties.

In the presence of plateaus, the outermost contour could contain more than one point with the same likelihood. When we replace one of the points in the plateau by a point with a greater likelihood, the volume cannot contract at all, since the outermost contour still contains other points at the same likelihood. Once we've replaced all the points in the plateau, the volume finally contracts. Crucially, however, it was not possible for any of the replacement points to affect the contraction, as the replacement points could never be the worst point whilst points in the plateau remain in the live points. The latter subtlety changes statistical estimates of the volume. Without plateaus, it is possible that a replacement point is or soon becomes the worst point, slowing the volume contraction. Without that possibility, the volume contracts faster and thus ordinary NS underestimates volume contraction and overestimates the evidence when there are plateaus.

\begin{figure*}
    \centering
    \includegraphics[width=0.7\textwidth]{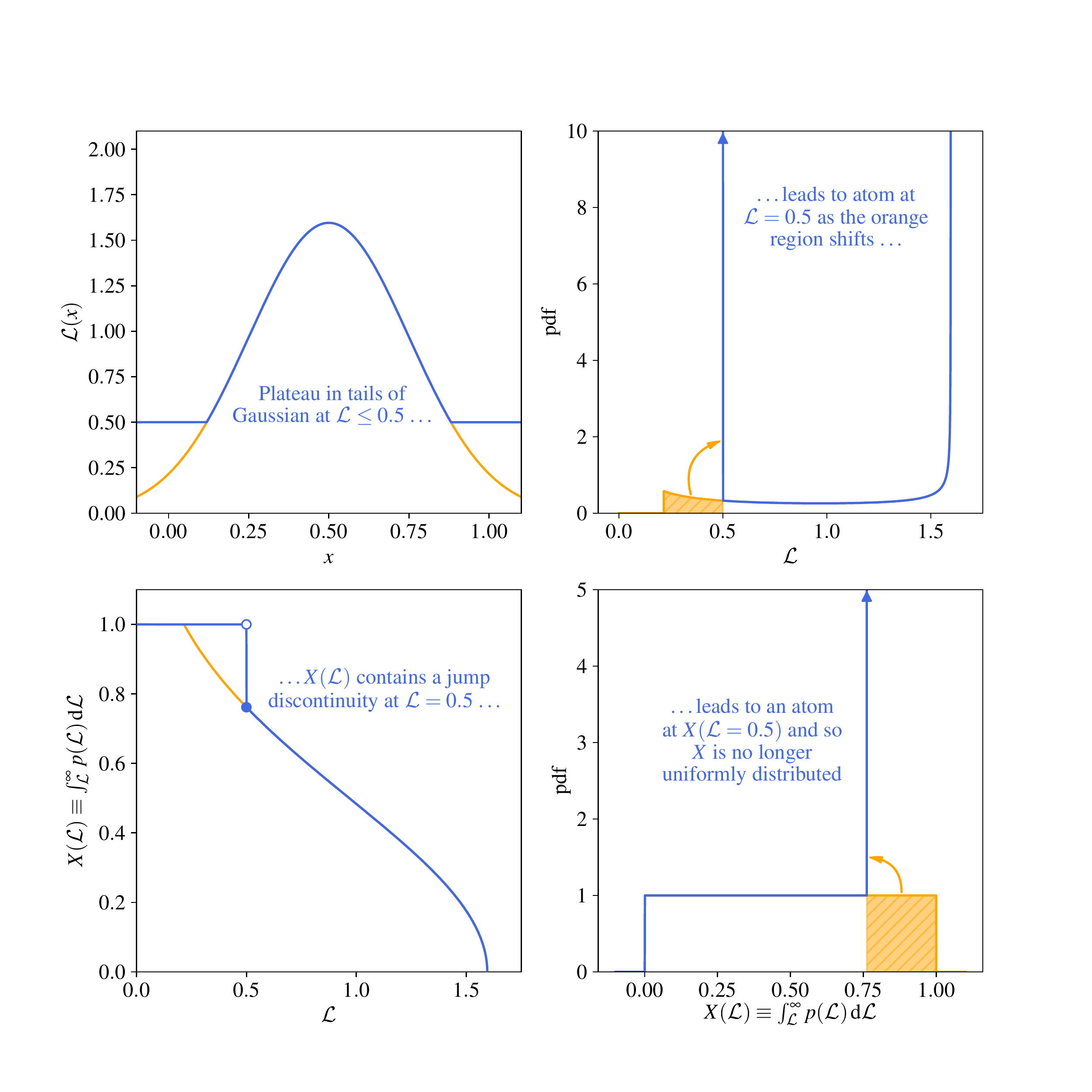}
    \caption{Infographic showing the impact of plateaus on assumptions in NS. We show an ordinary Gaussian (orange) and a  modified Gaussian with plateaus in the tails (blue). In the lower right panel, we see that the distribution of $X$ is no longer uniform, breaking assumptions in NS.}
    \label{fig:plateau_infographic}
\end{figure*}

The problem is illustrated by \cref{fig:plateau_infographic}. We show a Gaussian likelihood with mean $\mu = 0.5$ and standard deviation $\sigma = 0.25$ (orange) and a modified Gaussian likelihood with a plateau in its tails at $\plateaulike = 0.5$ (blue). In each case we consider a flat prior for the parameter $x$ from $0$ to $1$. The upper right panel shows the prior distribution of the likelihood. The plateau manifests as an atom in that distribution at \plateaulike. The  lower left panel shows the resulting enclosed prior volume as a function of the likelihood. The plateau causes a jump discontinuity at \plateaulike. Finally, the lower right panel shows the distribution of $X(\like)$. The plateau leads to an inaccessible region between about $0.8$ and $1$ and an atom of probability mass at $X(0.5) \approx 0.8$, and so $X$ is not uniformly distributed.

\begin{figure}
    \centering
    \includegraphics[width=0.8\linewidth]{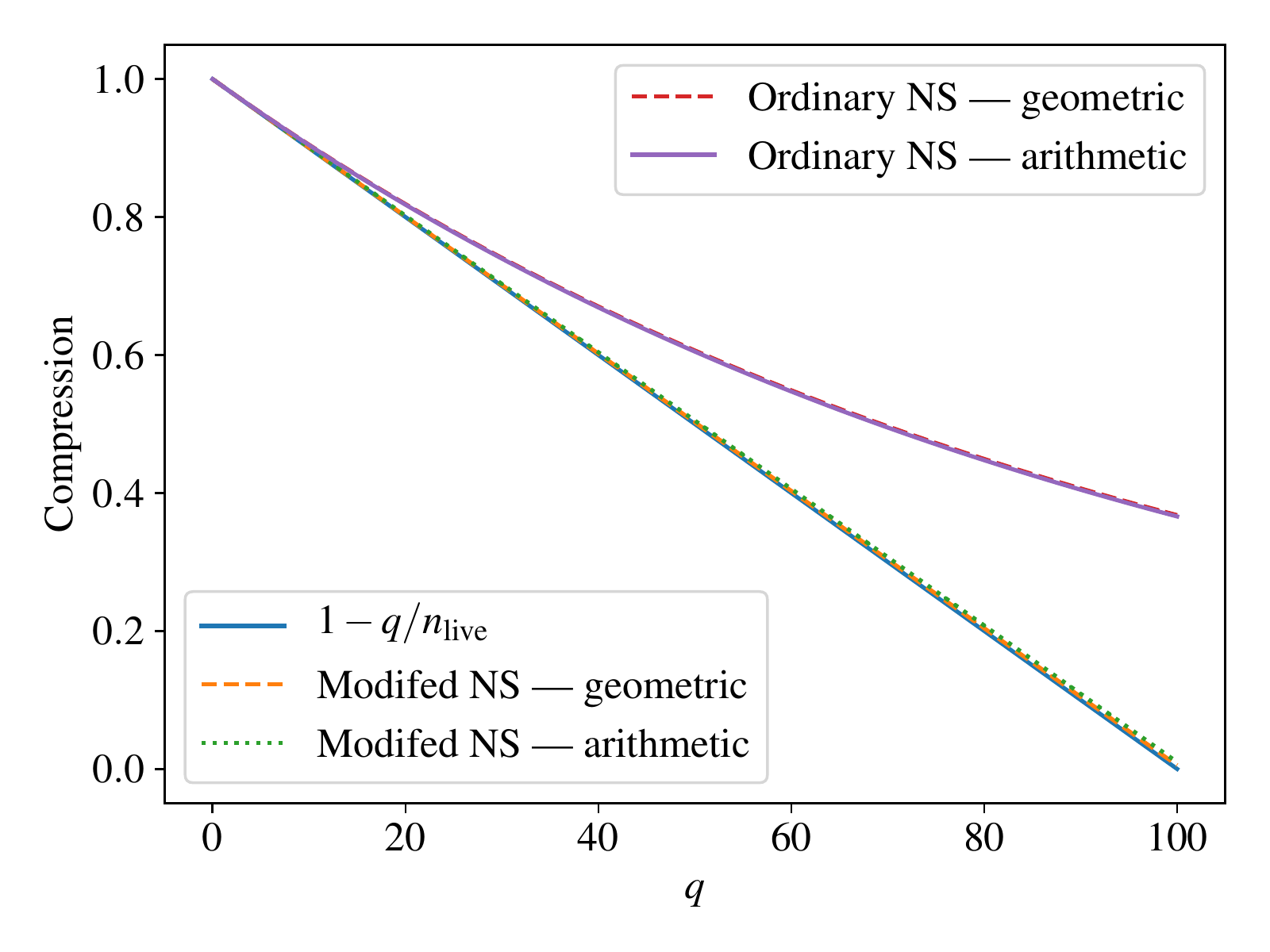}
    \caption{The compression from ordinary and modified NS when $q$ of 100 live points lie in a plateau. The correct linear compression is shown for reference (solid blue).}
    \label{fig:approximation}
\end{figure}

% The fact that $X$ is no longer uniformly distributed leads to faulty NS estimates of $X(\like)$ and thus to faulty estimates of posterior weights and the evidence.
Indeed, in ordinary NS, if $q$ outermost live points were in a plateau, we would compress by 
\begin{equation}\label{eq:ns_q_compression}
    e^{-\frac{q}{\nlive}}.
\end{equation}
We should, however, compress by about
\begin{equation}\label{eq:expected_compression}
   1 - \frac{q}{\nlive},
\end{equation}
which is an unbiased estimate of the volume outside the plateau based on binomial statistics (see \cref{sec:distributions} for further discussion). The compressions are only similar for $q \lesssim 0.5 \nlive$, since
\begin{equation}
    e^{-\frac{q}{\nlive}} \approx 1 - \frac{q}{\nlive} + \mathcal{O}\left(\frac{q^2}{\nlive^2}\right).
\end{equation}
The breakdown in the NS compression in \cref{eq:ns_q_compression} is shown in \cref{fig:approximation}. Note that this problem doesn't impact importance nested sampling \citep{Feroz:2013hea}, since it does not use the estimated volumes in \cref{eq:ns_q_compression}.

In fact, the arguments above show that in the presence of plateaus the inverse of $X(\lambda)$, denoted $\like(X)$ in overloaded notation, does not exist for all $ 0 \le X \le 1$ (see the lower right panel in \cref{fig:plateau_infographic}). As shown in \citet{2020arXiv200508602S}, in this case we should instead consider the generalised inverse
\begin{equation}
    \bar\like(X) \equiv \left\{\sup \lambda \in \image \like : X(\lambda) > X \right\},
\end{equation}
with which the evidence may be written as
\begin{equation}\label{eq:generalized}
    \Z = \int_0^1 \bar\like(X) dX.
\end{equation}
We now introduce a modified NS algorithm that correctly computes the evidence even in the presence of plateaus via \cref{eq:generalized}.  We summarise the treatment of plateaus in existing popular NS software in \cref{tab:codes}.

\begin{table*}
\begin{tabular}{llc}
\hline
Code & Handles plateaus & Definition of constrained prior\\
\hline
\code{nestle-0.2.0}~\citep{nestle} & No & $\like \ge \threshold$\\
\code{dynesty-1.0.0}~\citep{2020MNRAS.493.3132S} & No & $\like \ge \threshold$\\
\code{DIAMONDS-\href{https://github.com/EnricoCorsaro/DIAMONDS/commit/76409b22c9da782436b52e454a8b36bc78fca6f6}{\#76409b2}}~\citep{diamonds} & No & $\like \ge \threshold$\\
\code{MultiNest-3.1.2}~\citep{Feroz:2007kg,Feroz:2008xx,Feroz:2013hea}  & No but compatible with \anesthetic & $\like > \threshold$\\
\code{PolyChord-1.18.2}~\citep{Handley:2015fda,Handley:2015xxx} & No but compatible with \anesthetic & $\like > \threshold$\\
\code{DNest4-0.2.4}~\citep{2016arXiv160603757B} & Yes by $\uniform(0, 1)$ tie-breaking labels & $\like > \threshold$\\
Skilling's implementation~\citep{skilling_code} & Yes by user-specified tie-breaking labels & $\like > \threshold$\\
\hline
\end{tabular}
\caption{\label{tab:codes} Comparison of treatment of plateaus in popular NS software. The example program in Skilling's implementation specifies $\uniform(0, 1)$ tie-breaking labels.}
\end{table*}

\section{Modified NS algorithm}\label{sec:modified_ns}

In our modification to NS, we remove \emph{all} live points at the contour $\like = \threshold$ \emph{one by one without replacement}, contracting the volume after each removal. If there is a plateau, there may be more than one such point; if not, our algorithm reduces to ordinary NS. After removing all such points, we finally replenish the live points by adding points sampled from the prior subject to $\like > \threshold$, as usual. 

After removing a point, the number of live points drops by one, such that if we were to remove $i=1, \ldots, q$ such points (i.e., if there were $q$ points in the plateau) we would compress by
\begin{equation}\label{eq:arithmetic}
\begin{split}
\prod_{i=1}^q \expectation{t_i} &= \frac{\nlive}{\nlive + 1} \cdot \frac{\nlive - 1}{\nlive} \cdots \frac{\nlive - (q - 1)}{\nlive - (q - 2)}\\
                                    &= 1 - \frac{q}{\nlive + 1} 
\end{split}
\end{equation}
if using an arithmetic estimate of the compression, and by
\begin{align}\label{eq:geometric}
\sum_{i=1}^q \expectation{\ln t_i} &= -\sum_{i=1}^q \frac{1}{\nlive - (i - 1)}\\  % = H_{\nlive - q} - H_{\nlive}\\
                                   &\approx \ln\left(1 - \frac{q}{\nlive}\right) 
                                   % + \mathcal{O}\left(\frac{q}{\nlive (\nlive - q)}\right)
                                   \quad\text{for $\nlive \gg q$}\label{eq:approx_geometric}
\end{align}
if using a geometric one.
%, where $H_n$ is the $n$-th Harmonic number.
Thus we find in both cases that the compression from removing $q$ points would be about $1 - q / \nlive$ in agreement with \cref{eq:expected_compression}, the difference being of order $1 / \nlive$. The difference is noteworthy when $q \simeq \nlive$, in which case the unbiased estimate of the contraction would be zero, but the above estimates are about $1 / \nlive$.

% figure showing side by side algorithms
\SetAlCapNameFnt{\small}
\SetAlCapFnt{\small}

% for putting algorithms in minipage
\makeatletter
\newcommand{\removelatexerror}{\let\@latex@error\@gobble}
\makeatother

% skip line
\newcommand{\algoskipline}{
\DontPrintSemicolon%
\;%
\PrintSemicolon%
}

\definecolor{change}{rgb}{1.0, 0.01, 0.24}

\begin{figure*}
\begin{minipage}[t]{0.46\textwidth}%
\vspace{0pt}%
\begingroup%
\removelatexerror%
\begin{algorithm}[H]%
\SetAlgoLined
 Sample $\nlive$ points from the prior --- the live points\;
 Let $P$ be the set of live points\;
 Set $X_0 = 1$\;
 Initialise evidence, $\Z = 0$\;
 Initialise $i = 0$\;
 \While{$i \le \text{number of iterations}$}{
    Let $\threshold$ be the minimum $\like$ of the live points\;
    Let $r$ be the live point with $\like = \threshold$\;
    \algoskipline 
    Increment iteration, $i = i + 1$\;
    Contract volume, $X_i = X_{i -1} \cdot \exp\left(-1/\size(P)\right)$\;
    Assign importance weight, $w_i = (X_{i-1} - X_i) \cdot \threshold$\;
    Increment evidence, $\Z = \Z + w_i$\;
    Remove the point $r$ from the live points\;
    \algoskipline
    Add a new live point sampled from the prior subject to $\like > \threshold$\;
 }
 \KwRet{Estimate of evidence, $\Z$}
 \caption{Original NS.}
 \label{algo:original_ns}
\end{algorithm}
\endgroup
\end{minipage}%
\hspace{0.5cm}
\begin{minipage}[t]{0.46\textwidth}%
\vspace{0pt}%
\begingroup%
\removelatexerror%
\begin{algorithm}[H]%
\SetAlgoLined
 Sample $\nlive$ points from the prior --- the live points\;
 Let $P$ be the set of live points\;
 Set $X_0 = 1$\;
 Initialise evidence, $\Z = 0$\;
 Initialise $i = 0$\;
 \While{$i \le \text{number of iterations}$}{
    Let $\threshold$ be the minimum $\like$  of the live points\;
    Let \textcolor{change}{$R$ be the set of live points} with $\like = \threshold$\;
    \begingroup\color{change}
    \ForEach{point $r$ in $R$}{\color{black}
        Increment iteration, $i = i + 1$\;
        Contract volume, $X_i = X_{i -1} \cdot \exp\left(-1/\size(P)\right)$\;
        Assign importance weight, $w_i = (X_{i-1} - X_i) \cdot \threshold$\;
        Increment evidence, $\Z = \Z + w_i$\;
        Remove the point $r$ from the live points\;
        \color{change}
    }
    \endgroup
   Add \textcolor{change}{$n_\text{new} = \size(R)$ new live points} sampled from the prior subject to $\like > \threshold$\;
     
 }
 \KwRet{Estimate of evidence, $\Z$}
 \caption{Modified NS.}
 \label{algo:modified_ns}
\end{algorithm}%
\endgroup%
\end{minipage}
\caption{\label{fig:algorithms}\Cref{algo:original_ns}, original NS (left) and \cref{algo:modified_ns}, modified NS (right). The changes are shown in red.
If we were to add $n_\text{new} \neq \size(R)$ new live points in \cref{algo:modified_ns}, it would be a dynamic nested sampling algorithm~\citep{2019S&C....29..891H}.}
\end{figure*}

In \cref{algo:original_ns} and \cref{algo:modified_ns} we show the original and our modified NS, respectively. For concreteness we show the geometric estimator of the compression. We highlight the parts of the algorithm that are changed in red. The simple difference is that whereas in the original NS we replace a single live point, we instead replace \emph{all} the points sharing the minimum likelihood and contract the volume appropriately. If there are no plateaus, the algorithms are identical. 

This modified NS algorithm automatically deals with plateaus whenever they occur and can be applied retrospectively to NS runs. It automatically implements the robust NS algorithm of \citet{2020arXiv200508602S}, but with the advantage that it avoids the requirement that we explicitly decompose the evidence integral into a sum of contributions from plateaus and non-trivial contributions to be computed by NS. Other advantages are that our modification fits elegantly into ordinary NS, since the modifications are small, and that it is in the spirit of dynamic NS~\citep{2019S&C....29..891H}, since taking into account the changing number of live points is crucial.

We could, alternatively, evict all live points in the outermost contour, compress by $1 - q / \nlive$, and finally top up the live points. We don't favour this implementation of our idea, as it breaks the equivalent treatment of plateaus and non-plateaus in \cref{algo:modified_ns} and our use of ordinary NS compression factors, but it remains a valid possibility. This would use the same estimates of the plateau size as \citet{2020arXiv200508602S}. Lastly, we could alternatively remove plateaus entirely from the likelihood function by e.g., using the extended likelihood function in \cref{eq:extened_like}. If the tie-breaking terms, $\ell$, are random variates, they must be independent and identically distributed \citep[appendix B.5.4.3]{riley}. This, however, cannot be applied retrospectively to NS runs, and, as discussed in \citet{riley}, might lead to problems in specific NS implementations. 

Indeed, when using a random variate to break ties in a plateau, a new live point could lie anywhere in the plateau as for any point in the plateau we are equally likely to draw a tie-breaking random variate that leads to acceptance. Sampling from the constrained prior thus requires sampling from the whole plateau and contours above the plateau. This becomes unavoidably inefficient for large plateaus, as almost all tie-breaking random variates would lead to rejection. NS implementations that attempt to increase the sampling efficiency thus easily lead to faulty estimates of the evidence.

For example, in ellipsoidal rejection sampling, as used in \MN, one samples from ellipsoids that are constructed to enclose the current live points and with a volume similar to the current estimated volume, $X_i$, so that the ellipsoids typically shrink during an NS run. In the case of plateaus, this could easily lead to sampling from only a subregion of the whole plateau, as the ellipsoids could shrink as we compress through a plateau. In slice sampling, as used in \PC, we step out from a current live point to find the likelihood contour. To sample from the whole plateau, we should step out until at least the edge of the plateau. This, however, could fail, as the tie-breaking random variate when stepping out that far could by chance lead to rejection and thus we wouldn't step out far enough. These problems could possibly be avoided by elevating the tie-breaking random variate to a model parameter with a $\uniform(0, 1)$ prior. For ellipsoidal sampling in the presence of plateaus, this would allow the ellipsoids to contract only in that dimension and without encroaching on the plateau.

\subsection{Distribution of compression factor}\label{sec:distributions}

In our modified NS, we apply ordinary NS compression factors taking into account the dynamic number of live points. This assumes beta distributions for the compression factor. When dealing with plateaus, \cite{2020arXiv200508602S} consider estimating the compression using the fact that the number of points inside the plateau should follow a binomial distribution parameterized by the size of the plateau. Let us consider more carefully the differences in estimates of the compression factor from these two approaches. In the latter, the number of live points, $q$, that fall in the outermost contour plateau follows a binomial
\begin{equation}
    q \sim \BinomDist(\nlive, 1 - t) 
\end{equation}
where  $1 - t$ is the size of the plateau and thus $t$ is the compression factor. The probability mass function is thus
\begin{equation}
  \Pg{q}{t} \propto t^{\nlive - q} \, (1 - t)^q,
\end{equation}
for $q$ points in the plateau and $\nlive - q$ points above it. We could make inferences about $t$ by computing a posterior distribution,
\begin{equation}\label{eq:binom}
 \pg{t}{q} \propto \Pg{q}{t} \p{t} \propto t^{\nlive - q} \, (1 - t)^q \, \p{t},
\end{equation}
where $\p{t}$ is our prior for the size of the non-plateau region. For $ \p{t} = \text{const.}$ this is in fact the probability density for a $t \sim \BetaDist(\nlive + 1 - q, q + 1)$ distribution.

On the other hand, taking the approach in modified NS, the number of points that lie in the plateau $q$ is no longer treated as a random variable. Instead, the factor $t$ is assumed to follow a beta distribution when $q$ points are removed~\citep{2014AIPC.1636..100H},
\begin{equation}
t \sim \BetaDist(\nlive + 1 - q, q).
\end{equation}
% The beta distribution parameters are $\alpha = \nlive + 1 - q$ and $\beta = q$. 
The density for $t$ is
\begin{equation}\label{eq:beta}
  \pg{t}{q} \propto t^{\nlive - q} \, (1 - t)^{q - 1}
\end{equation}
corresponding to $q - 1$ points below $\threshold$, $\nlive - q$ points above it, and \emph{one point at $\threshold$.} 

The fact that we must consider $q - 1$ points below $\threshold$ and one point at $\threshold$, rather than $q$ points inside a plateau, results in a factor of $(1 - t)$ difference between \cref{eq:binom,eq:beta}. A further factor of $\p{t}$ originates from any prior knowledge about the size of the plateau. If the factors may be neglected inferences based on ordinary NS compression may be reliable. 

With a flat prior for the unknown size of the plateau, the difference is that between a $\BetaDist(\nlive + 1 - q, q)$ and a $\BetaDist(\nlive + 1 - q, q + 1)$ distribution. As shown in in \cref{fig:uncertainty}, the first two moments are similar, with only moderate differences of order $1 / \nlive$ even when $q \simeq 1$ and $q\simeq \nlive$.\footnote{The fact that the distributions as functions of $t$ are approximately identical is enough to ensure that our inferences are approximately identical~\citep{berger1988likelihood}. This holds despite the fact that in the binomial case the number of live points in the outermost plateau is a random variable and the size of the plateau $1 - t$ is fixed, and in the beta case the compression factor $t$ is a random variable and $q$ is fixed. This needn't be true for frequentist estimates of the factor $t$.}
The factor vanishes entirely for a logarithmic prior for the unknown size of the plateau,
\begin{equation}
    \p{t} \propto \frac{1}{1 - t} \quad\text{and so}\quad \p{f} \propto \frac1f.
\end{equation}
Thus we see that our modified NS treatment is approximately the same as the binomial treatment, the full details of which would depend on a prior for the size of the plateau. If the differences are important, though, we could instead use the binomial statistics, as discussed at the end of \cref{sec:modified_ns}, and a more careful treatment of the prior.

\subsection{Error estimates}\label{sec:errors}

The classic NS error estimate
\begin{equation}
    \Delta \log\Z \approx \sqrt{\frac{H}{\nlive}},
\end{equation}
where $H$ is the Shannon entropy between the posterior and prior, assumes the ordinary NS compression and a constant number of live points, and so is not applicable to our modified NS. The \anesthetic\ error estimates, however, already account for a dynamic number of live points, and so are applicable to the modified NS algorithm. The \anesthetic\ error estimates are found  \citep[as initially suggested by][]{Skilling:2006gxv} through simulations; sequences of possible compression factors are drawn from beta-distributions and used to make a set of estimates $\ln\Z$. One could alternatively compute an analytic estimate using the equivalents of the arithmetic expressions in \cref{eq:t} for the variance, which can be found in Appendix B of \citet{Handley:2015xxx}. Future versions of \texttt{anesthetic} will also support such analytic estimates. 

\begin{figure}
    \centering
    \includegraphics[width=0.8\linewidth]{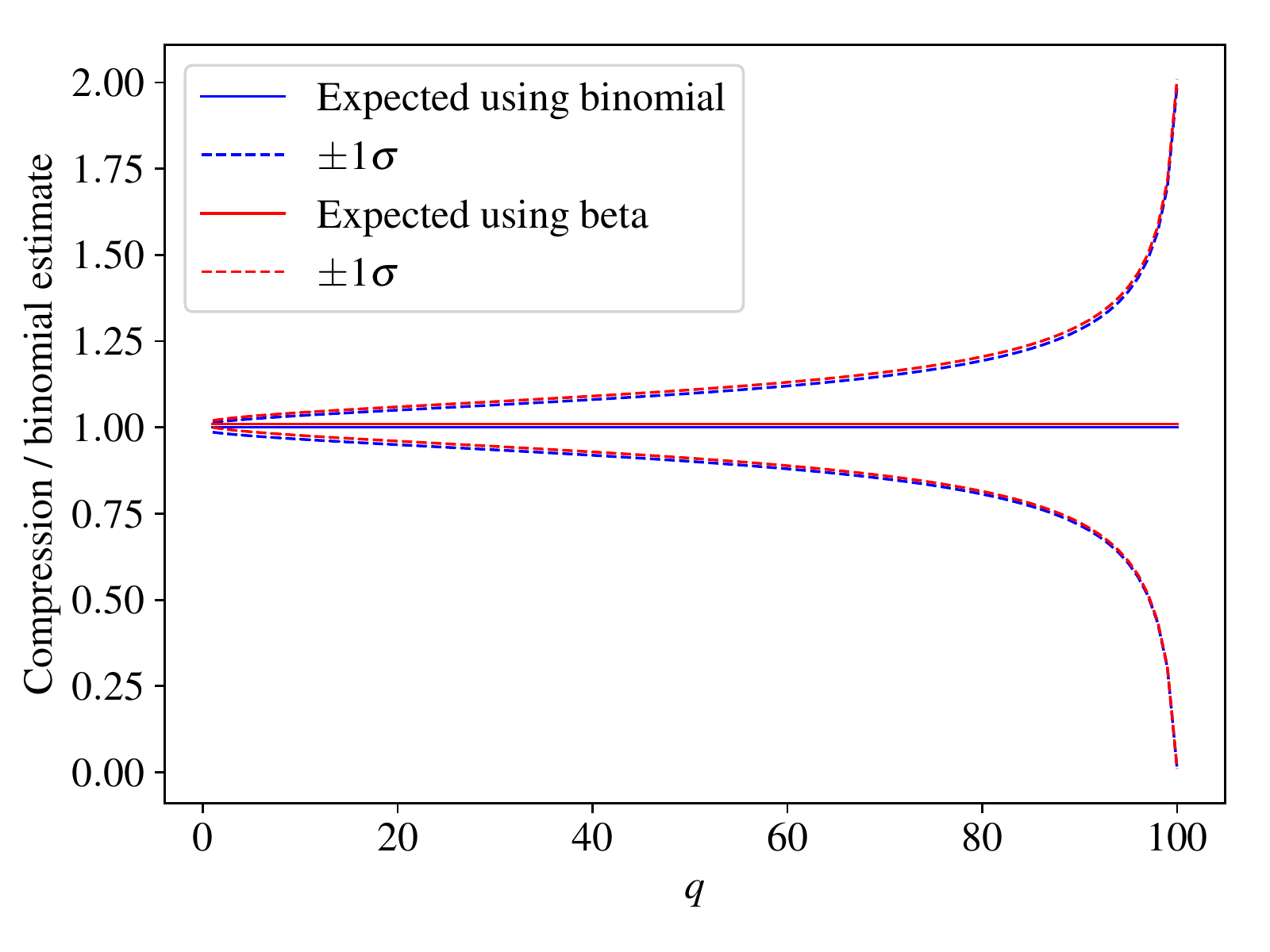}
    \caption{Estimates of compression for $\nlive = 100$ and $q$ points in the plateau from the binomial (blue) and beta (red). The estimates are shown relative to the ones from the binomial. The dashed lines show one standard deviation uncertainties.} 
    \label{fig:uncertainty}
\end{figure}

Clearly, however, compression estimates for plateaus suffer from increased uncertainties, since we dynamically reduce the number of live points during the plateau period of an NS run. In fact, if the plateau at $\plateaulike$ makes up a large fraction $f$ of the current prior volume, the error in the estimated compression could be substantial, as few points lie outside the plateau. Indeed, we show that the fractional error in the compression blows up when $q \simeq \nlive$ in \cref{fig:uncertainty}. In such cases, there may exist efficient schemes for dynamically increasing the number of live points to ensure that sufficient points lie above the plateau. For example, in \cref{algo:modified_ns} we must ultimately replenish the live points such that there are $\nlive$ live points at  $\like > \plateaulike$. We could instead dynamically increase the number of live points immediately prior to the plateau by sampling them from $\like \ge \plateaulike$, stopping at $\ndynamic$ once $\nlive$ of the $\ndynamic$ points lie at $\like > \plateaulike$. Once we remove points from the plateau, there would be $\nlive$ live points remaining, and no need to top up the live points.

In the worst case scenario, no live points land in contours above the plateau, e.g., the likelihood function shows a tiny peak above a broad plateau. At this point, it would be unclear whether to proceed and if so so how to do so efficiently, since the live points in the plateau provide no clues about the presence or location of the peak. This problem would affect all NS variants known to us.

% We could proceed by replacing the worst live point in the plateau using a random variate as a tie-breaker. We would eventually discover the peak, but samples drawn from the plateau that were rejected by the tie-breaker would be completely wasted. Our proposal to dynamically increase the number of live points until a sufficient number of live points lie in the peak would, on other other hand, ensure that all samples reduce uncertainty on the compression estimate, as samples become live points regardless of whether they lie in or above the plateau.

\section{Examples}\label{sec:examples}

We now consider a few examples. First, in \cref{sec:examples_simple} we apply our modified NS to examples considered in \citet{2020arXiv200508602S}. Second, in \cref{sec:wedding_cake} we construct a `wedding-cake' function that exhibits a series of plateaus and check that our modified NS correctly computes the evidence.

\subsection{Examples from \protect\citet{2020arXiv200508602S}}\label{sec:examples_simple}

\begin{figure}
    \centering
    \includegraphics[width=0.8\linewidth]{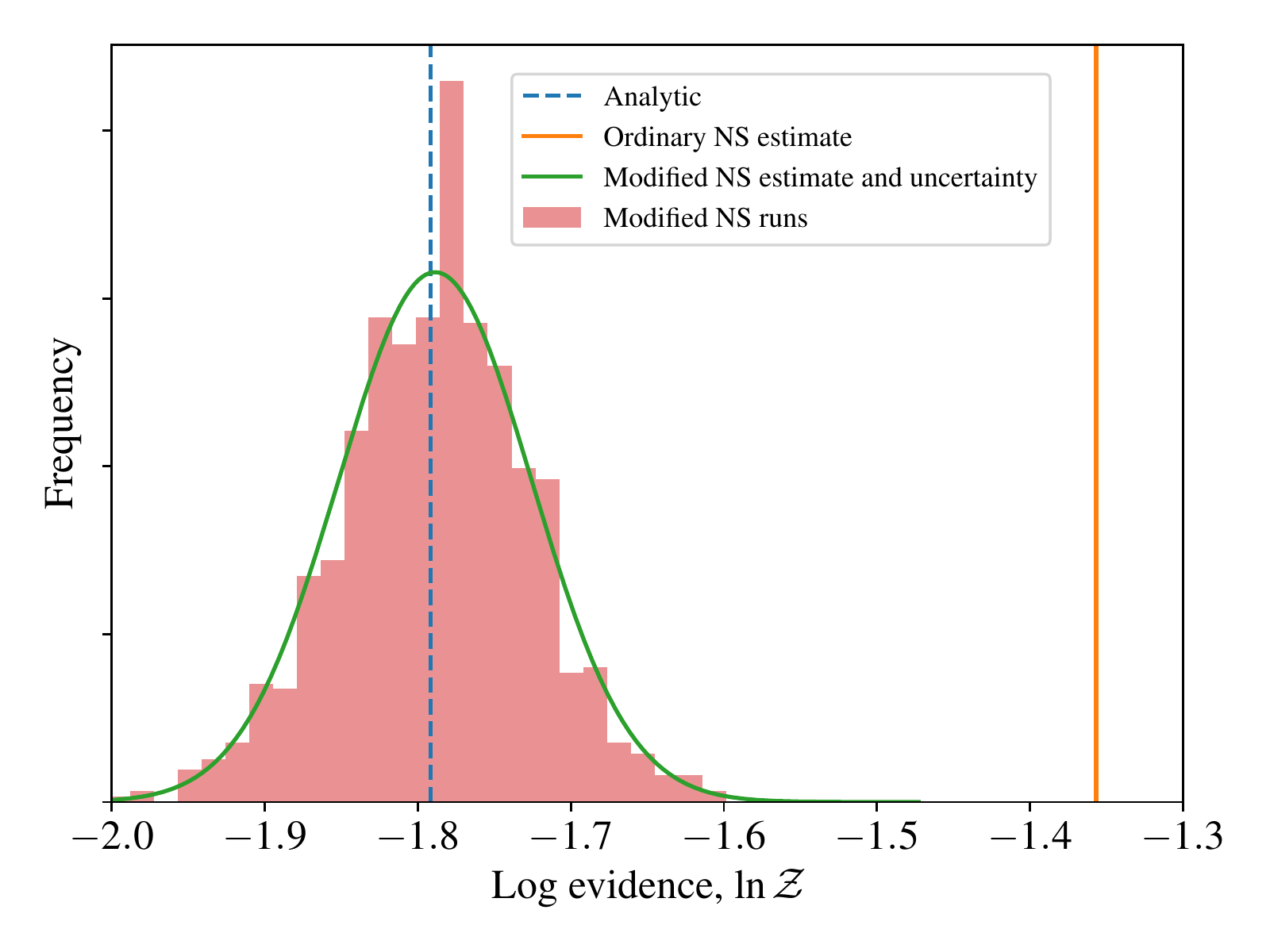}
    \caption{Repeated modified NS runs on Scenario 2 of \citet{2020arXiv200508602S}. We show the results from 1000 runs (red histogram) and the average mean and uncertainty estimate from the single modified NS runs as a Gaussian (green). For reference, we show the original NS result (orange) and the analytic result (dashed blue). }
    \label{fig:hist}
\end{figure}

For plateaus at the base of the likelihood function at $\like = 0$, we overestimate the evidence by a factor
\begin{equation}
    \Delta\log\Z = \log\left(\frac{e^{-f}}{1 - f}\right) = -\log(1 - f) - f
\end{equation}
where $f$ is the fraction of the prior volume occupied by the plateau. For example, in Scenario 2 of \citet{2020arXiv200508602S}, the plateau occupies $f = 2/3$ of the prior volume at the base of the likelihood, and we find $\Delta\log\Z \approx 0.432$, in good agreement with the difference found numerically in \citet{2020arXiv200508602S}, which was $0.433$. 
% -1.7916 - -1.3582 both numbers on p31 
In \cref{fig:hist}, we check the results of this problem with modified NS in 1000 repeat runs with 500 live points. We find that the histogram of $\ln\Z$ estimates form a Gaussian peak (red bars) around the analytic result (dashed blue). The spread was well-described by the average uncertainty estimates from single runs of modified NS (green). The original NS results (orange), on the other hand, lie well away from the analytic result, as expected.

For plateaus at the peak of the likelihood function at $\like = \max\like$, the error depends on the stopping conditions and treatment of final live points in NS. In e.g., the \MN implementation of NS, the run converges if all live points share the same likelihood, and the remaining evidence is correctly accounted for. If the run halts but the remaining evidence isn't accounted for, we underestimate the evidence by a term
\begin{align}
\Delta\Z &= f \max\like\\
\Delta\log\Z &= \log \left(\Z + \Delta\Z\right) - \log\Z
\end{align}
where $f$ is the fraction of the prior volume occupied by the plateau. For example, in Scenario 3 of \citet{2020arXiv200508602S}, the plateau occupies $f \simeq 0.161$ of the prior volume at the peak of the likelihood at $\max\log\like \approx -2.21$, and we find $\Delta\log\Z \approx 1.006$, in good agreement with the difference found numerically in \citet{2020arXiv200508602S}, which was $1.007$.
% -4.5861 - -3.5790 numbers on p33 and p34

\subsection{Wedding cake likelihood}\label{sec:wedding_cake}

\begin{figure}
   \centering
   \includegraphics[width=0.95\linewidth]{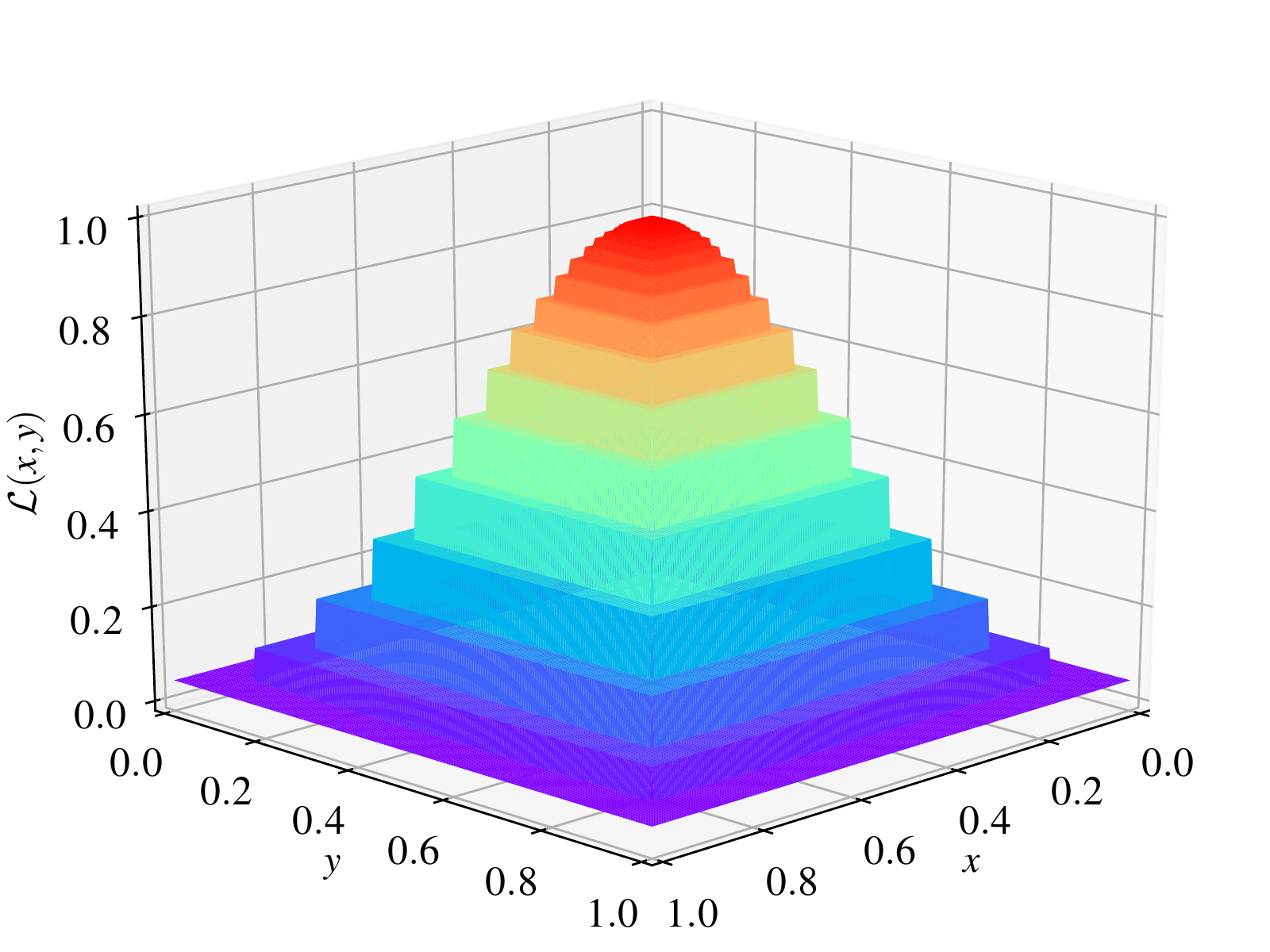}

   \includegraphics[width=0.95\linewidth]{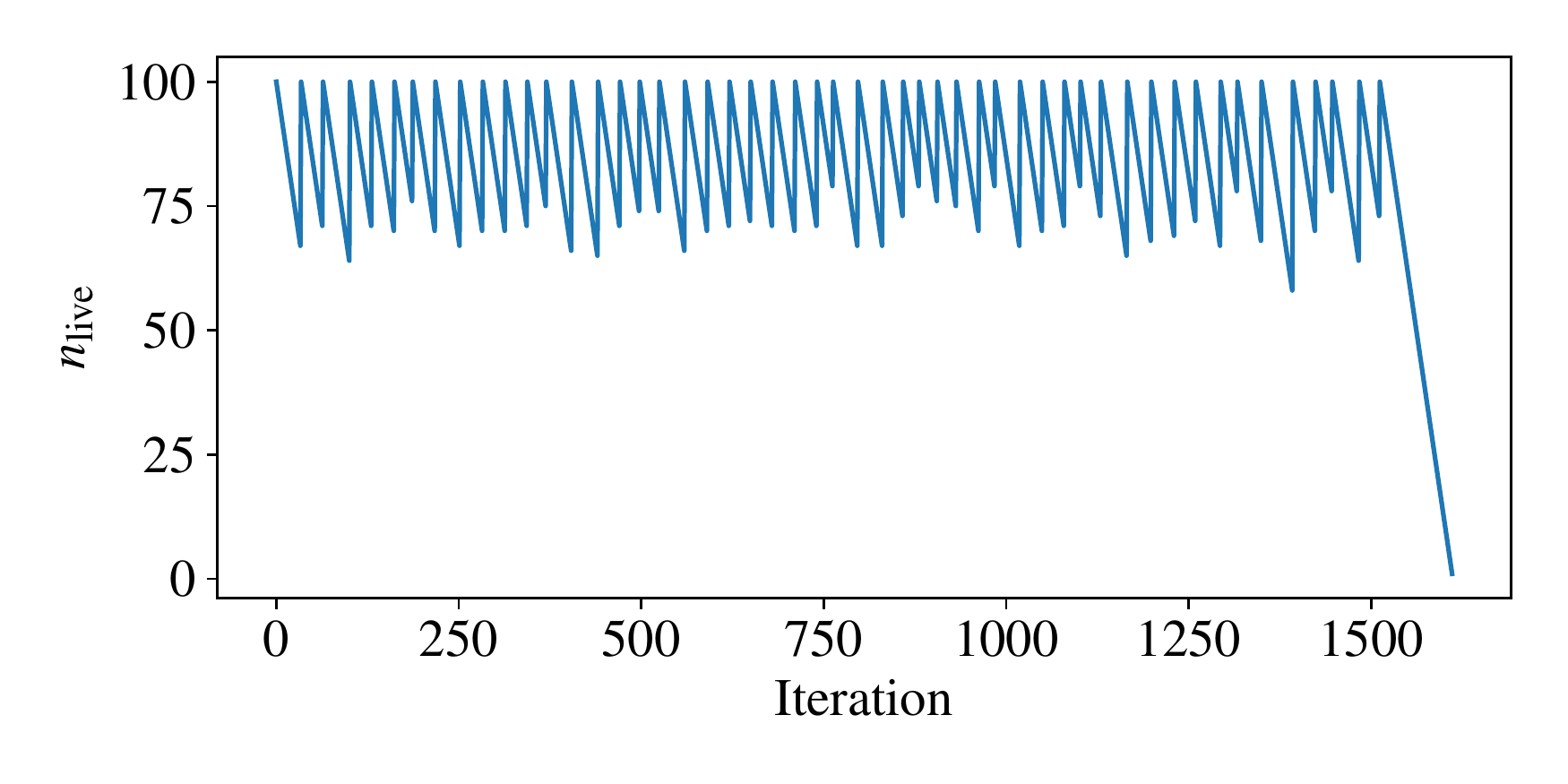}
   \caption{Top: Example of the wedding cake likelihood function in two dimensions for $\alpha = 0.7$ and $\sigma = 0.2$. Bottom: The number of live points used in the calculation of the evidence. The target value is $\nlive=100$, but in plateaus as points are discarded one-by-one without replacement this causes $\nlive$ to drop until the plateau is passed. The variability in the number of points discarded can be seen by the varying depth of each local minimum in the number of live points.}
    \label{fig:wedding_cake}
\end{figure}

We now construct a semi-analytical example for which we can numerically confirm our approach. Consider
an infinite sequence of concentric square plateaus of geometrically decreasing volume $\alpha^i (1-\alpha)$ for $i = 0, 1, 2, \ldots$. The edges of the plateaus lie at 
\begin{equation}
    r_i = \alpha^{i/D}/2 \quad\text{and}\quad r = \left|\boldsymbol{x} - 1/2\right|_\infty
\end{equation}
for $i = 0, 1, 2, \ldots$, where $|\boldsymbol{x}|_\infty \equiv \max_j(|x_j|)$ denotes the infinity norm. We define the height of each plateau to have a Gaussian profile:
\begin{equation}
    \log\like = -\sum_{i=0}^{\infty} \frac{r_i^2}{2\sigma^2}  \, \mathds{1}_{r_{i+1} < r \le r_{i}}
\end{equation}
where $\mathds{1}$ is an indicator function that, for any given $r$, selects a single term in the sum. The resulting likelihood is therefore a set of hypercuboids with a hypercubic base of side length $r_i$ and height $\exp(-{r_i^2}/{2\sigma^2})$. If the base is two-dimensional, this creates a tiered ``wedding cake'' surface, as can be seen in \cref{fig:wedding_cake}.
The $i$ selected by the indicator function is in fact,
\begin{equation}
    i(r) = \left\lfloor D\log_\alpha 2 r \right\rfloor
\end{equation}
where $\left\lfloor y \right\rfloor$ the floor function (namely the greatest integer less than or equal to $y$), 
enabling us to write
\begin{equation}
     \log\like = - \frac{\alpha^{2 i(r) / D} }{8\sigma^2}   
\end{equation}
Given that the volume of the region $[r_i < r < r_{i-1}]$ is $\alpha^{i}(1-\alpha)$, the evidence can be expressed as:
\begin{equation}
    \Z = \sum_{i=0}^\infty e^{-\alpha^{2i/D}/8\sigma^2} \alpha^i (1-\alpha) 
\end{equation}
which as an infinite series converges sufficiently rapidly and stably to be evaluated numerically for any number of dimensions $D$, but if speed is a requirement then a Laplace approximation shows that one only needs to consider the terms in the series around  
\begin{equation}
    i \sim\sqrt{\frac{D}{2}}\frac{\log (4 D \sigma^2) - 1 \pm \mathcal{O}(\text{a few})}{\log \alpha}.
\end{equation}

For reference, putting all of these equations together, the likelihood can be computed as:
\begin{equation}
    \log\like(\boldsymbol{x}) = - \frac{\alpha^{2  \left\lfloor D\log_\alpha 2 \left|\boldsymbol{x} - 1/2\right|_\infty  \right\rfloor / D} }{8\sigma^2}   
\end{equation}
where $\alpha$ is a hyperparameter controlling the depth of the plateaus, and $\sigma$ controls the width of the overall Gaussian profile.

This likelihood forms part of the \texttt{anesthetic} test suite, which confirms that the approach suggested in this paper recovers the true likelihood.

The wedding cake likelihood can be very useful for testing nested sampling implementations as unlike a traditional Gaussian it can be trivially sampled from using a simple random number generator, has no unexpected edge effects as the boundaries of the prior are also a likelihood contour.

\section{Conclusions}\label{sec:conclusions}

Following from \citet{riley,2020arXiv200508602S}, which showed formally why NS breaks down if there are plateaus in the likelihood, we first presented the problem of plateaus in an informal but accessible way. We then constructed a modified version of the NS algorithm. The simple modification permits it to remove points in a plateau one by one, without replacement.  Once all points in a plateau are removed, the live points are finally replenished. This leads to correct compression in the presence of plateaus and ordinary NS in their absence.

We discussed examples from \citet{2020arXiv200508602S}, shedding light on them by finding the impact of plateaus on ordinary NS in simple analytic formulae. The impact was previously shown only numerically. Lastly, our especially constructed wedding-cake problem showed a case with multiple plateaus. The modified NS algorithm successfully dealt with them.

Runs from popular NS software such as \PC and \MN may be resummed retrospectively via the modified NS algorithm using \code{anesthetic} starting from version \version. Our modified NS makes a minimal change to ordinary NS, to the extent that we recommend it becomes the canonical version of NS.

\section*{Acknowledgements}

AF was supported by an NSFC Research Fund for International Young Scientists grant 11950410509.
WH was supported by a George Southgate visiting fellowship grant from the University of Adelaide, Gonville \& Caius College, and STFC IPS grant number ST/T001054/1. We thank Brendon Brewer for valuable comments and discussion on the manuscript.

\section*{Data availability}
There is little data associated with this paper, though any data or code will be shared on reasonable request to the corresponding author.

\bibliographystyle{mnras}
\bibliography{references}

\appendix

% Don't change these lines
\bsp	% typesetting comment
\label{lastpage}
\end{document}